\documentclass[10pt,letterpaper,twocolumn]{article}

\usepackage[letterpaper,       % 彻底抛弃 A4，切换为美国标准 Letter 纸张
            top=1.91cm,        % 官方定义：0.75 英寸顶部留白
            bottom=2.54cm,     % 官方定义：1.00 英寸底部留白（给页脚留足空间）
            left=1.91cm,       % 官方定义：0.75 英寸左边距
            right=1.91cm,      % 官方定义：0.75 英寸右边距
            columnsep=0.84cm,  % 官方定义：0.33 英寸标准双栏间距（中缝）
            headheight=14pt]{geometry}
            
\usepackage[T1]{fontenc}
\usepackage[utf8]{inputenc}
\usepackage{lmodern}
\usepackage{microtype}
\usepackage{etoolbox}
\patchcmd{\thebibliography}{\list}{\footnotesize\list}{}{}

\usepackage[sb]{libertine} %
\usepackage[varqu,varl]{inconsolata} %
\usepackage{amsthm}
\usepackage[libertine]{newtxmath} %

\ifdefined\pdfmapfile
  \pdfmapfile{+libertine.map}
  \pdfmapfile{+zi4.map}
  \pdfmapfile{+newtx.map}
\fi

\usepackage{fancyhdr}
\renewcommand{\headrulewidth}{0pt} % 默认去掉后续页面的页眉横线

\usepackage{afterpage}
\usepackage{xcolor}
\definecolor{ttorange}{HTML}{FF5000}      %
\definecolor{ttorangeD}{HTML}{FF5000}     %

\definecolor{ttaccent}{HTML}{FFD54F}
\definecolor{ttink}{HTML}{1A1A1A}
\definecolor{ttgray}{HTML}{5A5A5A}
\definecolor{ttbg}{HTML}{FFF6EC}
\definecolor{ttline}{HTML}{EAD9C6}

\usepackage{graphicx}
\graphicspath{{./}{./assets/}}

\usepackage{amsmath,amsthm}
\usepackage{booktabs}
\usepackage{multirow}
\usepackage{caption}
\theoremstyle{plain}

\theoremstyle{definition}

\theoremstyle{remark}

\usepackage{enumitem}
\setlist[itemize]{leftmargin=1.2em,itemsep=1pt,topsep=2pt}
\setlist[enumerate]{leftmargin=1.4em,itemsep=1pt,topsep=2pt}

\usepackage{titlesec}
\titleformat{\section}
  {\normalfont\large\bfseries\color{ttink}}
  {\textcolor{ttorange}{\thesection}}{0.55em}{}
\titleformat{\subsection}
  {\normalfont\normalsize\bfseries\color{ttink}}
  {\textcolor{ttorange}{\thesubsection}}{0.55em}{}
\titleformat{\subsubsection}
  {\normalfont\normalsize\itshape\color{ttink}}
  {\thesubsubsection}{0.5em}{}
\titlespacing*{\section}{0pt}{9pt plus 2pt minus 1pt}{6pt}
\titlespacing*{\subsection}{0pt}{7pt plus 1pt minus 1pt}{3pt}

\usepackage{fancyhdr}
\renewcommand{\headrulewidth}{0.4pt}

\renewcommand{\headrule}{\hbox to\headwidth{\color{ttline}\leaders\hrule height \headrulewidth\hfill}}
\usepackage{fancyhdr}
\newbool{isfirstpage}
\booltrue{isfirstpage}

\renewcommand{\headrule}{%
  \ifbool{isfirstpage}{%
    \hbox to\headwidth{\color{ttline}\leaders\hrule height 0.4pt\hfill}%
  }{%
  }%
}

\usepackage[colorlinks=true,linkcolor=ttorange,citecolor=ttorange,
            urlcolor=ttorange]{hyperref}

\usepackage{tcolorbox}

\usepackage{xcolor,array,booktabs,colortbl}
\definecolor{egblue}{rgb}{0.937,0.965,0.996}
\newcommand{\egfirst}[1]{\multicolumn{1}{@{}>{\columncolor{egblue}[0pt][\tabcolsep]}l}{#1}}
\makeatletter
\newcommand{\egtightrows}{\noalign{\global\setbox\@arstrutbox\hbox{\vrule\@height.75\ht\@arstrutbox\@depth.75\dp\@arstrutbox\@width\z@}}}
\newcommand{\egnormalrows}{\noalign{\global\setbox\@arstrutbox\hbox{\vrule\@height\arraystretch\ht\strutbox\@depth\arraystretch\dp\strutbox\@width\z@}}}
\makeatother
\newcommand{\egnote}[1]{\multicolumn{12}{@{\hspace{6pt}}>{\raggedright\arraybackslash\tiny}p{0.5\textwidth}@{}}{#1}}

\usepackage{makecell,tabularx}
\providecommand{\Description}[1]{}
\hypersetup{pdftitle={E-GenRM: Generative Reward Modeling for Open-Ended E-Commerce Shopping Agents},
  pdfauthor={Ying Sun, Qiuyu Zhao, Xiao Wang, Xiyu Zhao, Gege Shi, Zenghui Sun, Jinsong Lan, Haihong Tang, Bo Zheng}}
\newcommand{\reporttitle}[1]{\def\thereporttitle{#1}}

\newcommand{\reportauthors}[1]{\def\thereportauthors{#1}}

\newcommand{\authormark}[2]{{\small #1}\textsuperscript{\color{ttgray}#2}}

\begin{document}

\reporttitle{E-GenRM: Generative Reward Modeling for Open-Ended E-Commerce Shopping Agents}

\reportauthors{%
  \authormark{Ying Sun}{1,*}, \authormark{Qiuyu Zhao}{1,*},
  \authormark{Xiao Wang}{2}, \authormark{Xiyu Zhao}{1},
  \authormark{Gege Shi}{1},
  \authormark{Zenghui Sun}{1}, \authormark{Jinsong Lan}{1},
  \authormark{Haihong Tang}{1}, \authormark{Bo Zheng}{1,\dag}\par
  \vspace{6pt}
  \small\textbf{{\textsuperscript{1}Taobao \& Tmall Group of Alibaba}} \quad
  \small\textbf{{\textsuperscript{2}Technische Universit\"at Wien}}\par
  \vspace{4pt}
  {\small\color{ttgray}*Equal contribution\quad
   {\dag}Corresponding author\par}
}

\twocolumn[{%
  \begin{center}
  \vspace*{10pt}
    {\huge\bfseries\color{ttink}\thereporttitle\par}
    \vspace{16pt}
    \thereportauthors
    \vspace{8pt}
  \end{center}
  \begin{center}
  \begin{tcolorbox}[
    width=0.92\textwidth,
    colframe=ttorange,
    colback=ttorange!3!white,
    arc=2mm,
    boxrule=0.8pt,
    left=35pt, right=35pt,
    top=22pt, bottom=22pt,
    boxsep=0pt,
  ]
    {\bfseries\color{ttorange}
    \centerline{\large\textbf{\textcolor{ttorange}{Abstract}}}}\par
    \vspace{8pt}
    {

We present \textbf{E-GenRM} (\textbf{E}-commerce \textbf{Gen}erative
\textbf{R}eward \textbf{M}odel), a generative reward model tailored to
e-commerce shopping agents. User expectations are shifting from simple
product queries to open-ended shopping tasks. However, existing trajectories and reward
designs remain largely bounded by current agent usage patterns and
known failures, providing limited guidance for further
capability improvement. These limitations call for an integrated reward modeling framework for open-ended shopping agent tasks. 
On the data side, E-GenRM leverages naturally occurring user behavior
across the shopping app to construct diverse, long-horizon conversational
trajectories beyond established usage patterns. On the modeling side,
E-GenRM first learns from behavior-grounded evaluation chains through
supervised fine-tuning. It is then optimized through criterion-guided
reinforcement learning, which rewards correct preference judgments and
evidence-supported factual claims under each criterion. Personalized
hard-negative training further improves its sensitivity to user preferences
reflected in shopping behavior, beyond explicit requirements. We additionally develop an active
evidence retrieval extension that acquires missing e-commerce information for
evaluation. On E-GenRM-Bench, E-GenRM-27B achieves 87.40\% preference
accuracy, outperforming the strongest baseline by 6.97 percentage points.
In a 14-day online A/B test on Taobao, a shopping agent post-trained with E-GenRM
improves the effective shopping-assistance rate by 1.39 percentage points
and user satisfaction by 2.33 percentage points.
}\par
    \vspace{12pt}
    {\footnotesize\color{ttgray}%
     \textbf{Date:} ~November 04, 2026 \par
     \vspace{3pt}
     \par
     \vspace{3pt}
     \hypersetup{urlcolor=ttgray}
     \textbf{Correspondence:}~\href{mailto:yudan.sy@taobao.com}{yudan.sy@taobao.com}\textsuperscript{*};
     \href{mailto:zhaoqiuyu.zqy@taobao.com}{zhaoqiuyu.zqy@taobao.com}\textsuperscript{*};
     \href{mailto:bozheng@alibaba-inc.com}{bozheng@alibaba-inc.com}\textsuperscript{\dag}\par}
  \end{tcolorbox}
  \end{center}
  \vspace{10pt}

}]

\section{Introduction}
\label{sec:introduction}

The effectiveness of reward-guided post-training hinges critically on the
quality and verifiability of the reward signal. This relationship has been
shown by reinforcement learning with verifiable rewards (RLVR)~\cite{lambert2025tulu3pushingfrontiers},
which has yielded substantial gains in mathematics and coding agents, where
correctness can be evaluated using well-defined rules or executable
feedback. However, shopping lacks similarly tractable rewards, as evaluation depends on dynamic user intent and heterogeneous product evidence
~\cite{intenttune,yeetal2025productagent,wang2026shoppingbench,tang2026revision}.
This complexity makes reliable reward modeling, and hence reward-guided post-training challenging for shopping agents.

As AI capabilities advance and shopping agents gain wider adoption, user expectations for these agents continue to rise. Traditional e-commerce search focuses on matching user queries with products, with performance typically measured by query--item relevance \cite{reddy2022shoppingqueries}. In practice, users must iteratively refine their intent through repeated queries, filters, and product browsing \cite{yeetal2025productagent}. Shopping agents offer a fundamentally new interaction paradigm. An ideal shopping agent should interpret complex, open-ended shopping goals and directly follow user instructions to retrieve fine-grained product information or suitable items, reducing the cognitive burden of interpreting and refining search results \cite{yao2022webshop,yeetal2025productagent,wang2026shoppingbench}.

This multi-turn interaction paradigm also shifts reward modeling from pointwise relevance to trajectory-level correctness. Relevance to the latest query alone is insufficient, as a response may overlook constraints established earlier in the interaction or recommend products unsupported by reliable evidence. Success must therefore be assessed against requirements accumulated throughout the trajectory, with each judgment grounded in reliable product evidence \cite{du2026ecomagentbench,yu2026shoppingcompanion}.
Learning a reward model that meets this standard requires shopping trajectories that expose diverse accumulated requirements and learning signals that provide feedback on individual requirements and decisions. Existing trajectories and reward designs are limited in both respects, which gives rise to two challenges.

\begin{figure}[t]
\centering
\includegraphics[width=0.98\columnwidth]{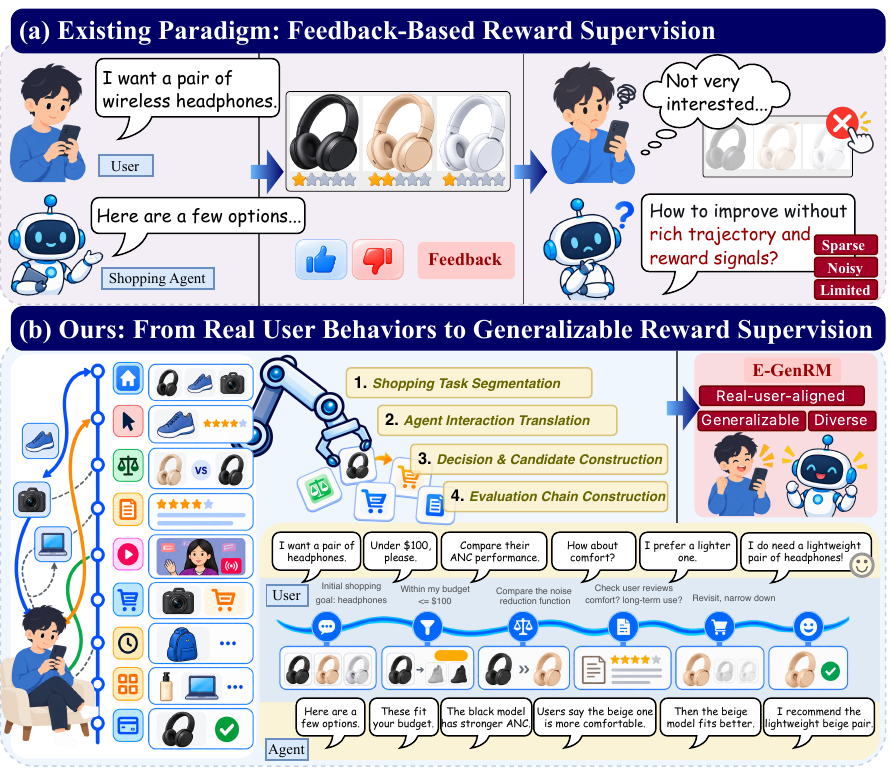}
\caption{From agent feedback to behavior-grounded reward supervision. (a) Existing methods rely on sparse and limited agent feedback. (b) E-GenRM converts real user behavior into evaluation chains, capturing diverse reward-relevant factors.}
\label{fig:data-construction}
\end{figure}

\afterpage{\global\boolfalse{isfirstpage}}
\noindent\textbf{Challenge 1: The Capability--Trajectory Feedback Loop.}
Reliable reward modeling requires diverse trajectories that expose behaviors
beyond current shopping-agent usage patterns. However, deployed trajectories
are endogenous to agent capabilities: users adapt their requests to supported
tasks, causing logged interactions to concentrate on established use cases and
known failures. Reward-model improvement is therefore constrained by
trajectories generated within existing capability boundaries (Figure~\ref{fig:data-construction}a).

\noindent\textbf{Challenge 2: Reward Aggregation and Requirement-Level Credit
Assignment.}
Shopping tasks involve interdependent decisions over long trajectories, with
reward signals often sparse and delayed. Shopping requirements (such as budget, functionality, and specifications) carry different weights across tasks, which fixed aggregation cannot capture. 
Scalar rewards further compress per-requirement judgments into a single score, diluting the feedback on each.
E-commerce platforms often list several products that meet the user's stated
requirements and differ only in unstated preferences, and coarse rewards cannot distinguish among them.
This information loss obscures which decisions drive success or failure,
providing limited guidance for policy improvement.

To address these challenges, we propose \textbf{E-GenRM}, a generative reward
model purpose-built for e-commerce. Leveraging high-value shopping
trajectories, E-GenRM provides expressive reward signals for post-training shopping agents.
Specifically, we collect more than 120
types of user behavior over one month and organize these heterogeneous signals into
semantically structured, task-specific interaction histories. As they originate from app-wide shopping activity rather than solely from
agent-mediated interactions, these histories capture diverse real-world
shopping patterns that can be translated into shopping-agent evaluation
trajectories, providing broader, more open-ended supervision beyond current
agent usage patterns. A teacher model performs this translation by deriving
task-specific criteria from the histories and constructing evidence-grounded
evaluation chains for candidate responses (Figure~\ref{fig:data-construction}b).

\begin{figure*}[!t]
\centering
\includegraphics[width=1.0\textwidth]{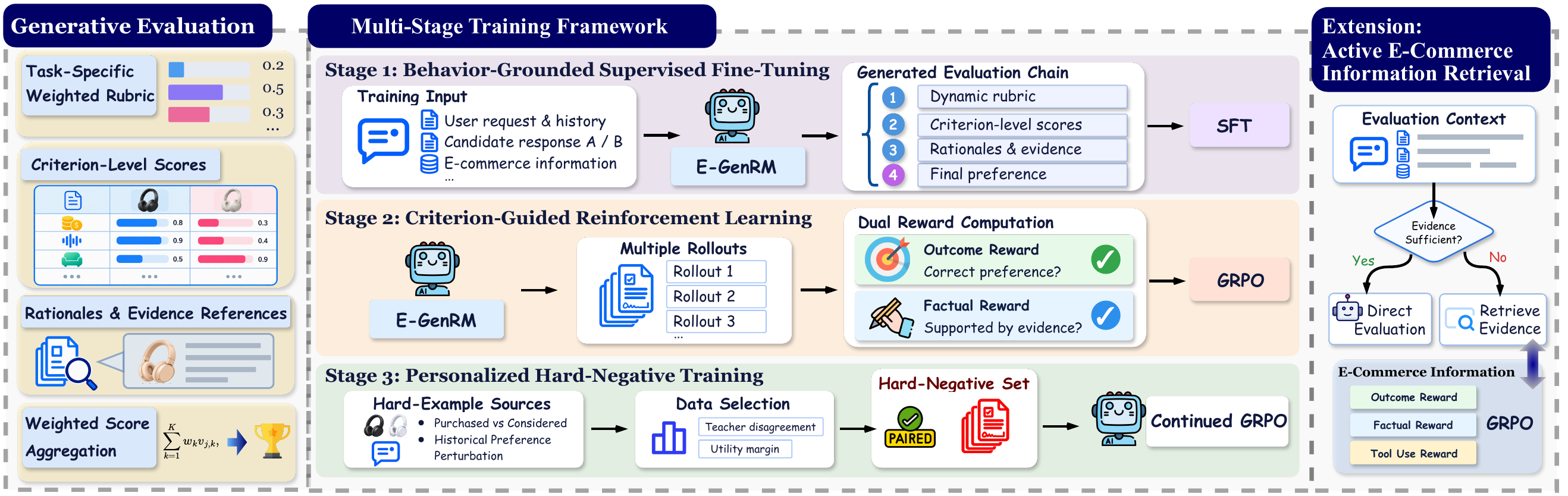}
\caption{Overview of E-GenRM, including the evaluation chain, three-stage training, and the active e-commerce information retrieval extension.}
\Description{The three-stage E-GenRM pipeline constructs evaluation chains from shopping behavior, optimizes outcome and factual criterion rewards, and focuses on close personalized comparisons. A separate extension acquires missing e-commerce information before applying the same evaluation chain.}
\label{fig:egenrm-framework}
\end{figure*}

{\emergencystretch=1.5em Using these evaluation chains, we train E-GenRM through a three-stage
framework comprising supervised fine-tuning, criterion-guided reinforcement
learning, and targeted hard-negative training (Figure~\ref{fig:egenrm-framework}). First, E-GenRM
learns to derive a task-specific rubric of weighted criteria from behavioral
histories and apply it to evidence-grounded evaluation. Second, criterion-guided reinforcement
learning rewards correct preference judgments and verifies
criterion-level factual claims against the visible evidence. Rather than directly
predicting a scalar score, E-GenRM produces evidence-grounded judgments for
individual criteria before deriving an overall preference. Factual feedback
on these judgments provides partial credit and denser supervision beyond
final preference correctness. In the final stage, personalized hard-negative
training emphasizes candidates that satisfy explicit requirements but
differ in behavior-supported preferences, improving the model's sensitivity
to personalized trade-offs.\par}

Beyond the three-stage training framework, we develop a separately trained
active e-commerce information retrieval extension to address cases where decisive
information is absent from the evaluator's context, as often occurs for long-tail
products or recently updated information. Building on tool-augmented reward
modeling~\cite{armthinker}, its tool-use objective aligns retrieval decisions
with whether the context lacks decisive evidence. The retrieved evidence is then
incorporated into the same rubric-based evaluation process, expanding
information access without altering the core judgment procedure. Our contributions are twofold.

{\emergencystretch=1.5em
\begin{itemize}
    \item We propose \textbf{E-GenRM}, which derives task-specific evaluation criteria from naturally occurring user behavior and translates diverse real-world shopping patterns into agent-evaluation trajectories, providing open-ended supervision for robust reward modeling. E-GenRM learns context-dependent, evidence-grounded evaluation through supervised fine-tuning, criterion-guided reinforcement learning, and personalized hard-negative training, with an active evidence retrieval extension that acquires missing e-commerce information when needed.
    \item We construct E-GenRM-Bench from real shopping behavior to evaluate
    reward models. E-GenRM-27B achieves 87.40\% preference accuracy on this
    benchmark, exceeding the strongest baseline by 6.97 percentage points. Using E-GenRM for shopping-agent post-training further improves
    effective shopping assistance and user satisfaction.
\end{itemize}\par}%

\section{Related Work}
\label{sec:related-work}

\subsection{E-Commerce Shopping Agents}

Shopping agents assist users with multi-step tasks through interaction. WebShop enables agents to search, navigate, and select products from user instructions \cite{yao2022webshop}, ProductAgent combines dialogue and retrieval to clarify ambiguous demands \cite{yeetal2025productagent}. ShoppingBench expands the task space to multi-product goals involving budgets and vouchers \cite{wang2026shoppingbench}. More recent benchmarks examine long-horizon interactions: EComAgentBench distributes intent across queries, profiles, and clarification exchanges \cite{du2026ecomagentbench}, while Shopping Companion studies cross-session preference inference and product-attribute verification \cite{yu2026shoppingcompanion}. These studies highlight the importance of interaction history and product evidence in shopping-agent evaluation.

User behavior has also been leveraged to improve agent capabilities. IntentTune uses behavioral signals to resolve underspecified intents \cite{intenttune}; LOFA combines purchase feedback with conversational instructions \cite{lofa}; and MACS integrates preference tracking with deterministic catalog filtering for reliable execution \cite{macs}. Despite this progress, existing shopping agents remain centered on localized capabilities such as product retrieval, intent understanding, and preference tracking, and are commonly evaluated by benchmarks designed around these capabilities. They remain insufficient to address users' open-ended and evolving shopping needs.

\subsection{Generative Reward Models}

Generative reward models formulate evaluation as text generation, allowing reasoning to precede scores or preferences. GenRM trains verifiers via next-token prediction and supports synthetic verification rationales \cite{zhang2025generativeverifiers}. RM-R1 generates sample-specific rubrics or reference solutions and combines reasoning distillation with reinforcement learning \cite{rmr1}. Reward Reasoning Models acquire reward reasoning through RL without supervised reasoning traces \cite{rewardreasoningmodel}. RubricRM extends rubric-based evaluation to image generation and editing using input-specific weighted criteria and dimension-level rewards \cite{rubricrm}. Together, these methods establish explicit evaluation chains as an alternative to opaque scalar scores.

Personalization and evidence acquisition address two complementary
requirements of evaluation. P-GenRM derives personas and scoring rubrics
from preference signals and uses user prototypes for test-time scaling~\cite{zhang2026pgenrm}. A separate line of work equips reward models with
evidence-seeking capabilities. ARM-Thinker learns to use tools to verify visual details and retrieve supporting documents \cite{armthinker}, but focuses on evidence grounding rather than recovering evaluation criteria from behavioral histories. Despite recent progress, generative reward modeling remains particularly challenging in shopping due to the complexity of the underlying data, decision rules, and supporting evidence.

\section{Problem Formulation}
\label{sec:problem-formulation}

We formulate the evaluation of shopping agent responses in open-ended shopping as context-dependent pairwise reward modeling.

A shopping trajectory contains multiple decision steps at which the agent responds to the user, such as product comparison, product information queries, and the final recommendation. At decision step $t$, the interaction history $h_t=(u_1,\ldots,u_t)$ ends with the current user request. Each $u_i$ records a user request or prior agent response in natural language. 

The evaluator receives context $x_t=(h_t,e_t)$ and candidate responses $y_t^A,y_t^B$, where $e_t$ is optional e-commerce information for the current task. The goal is to identify the response that better satisfies the user's current requirements and preferences.

The evaluator $R_\theta$ generates an evaluation chain $\tau_t$:
\begin{equation}
    \begin{aligned}
        \tau_t &= \bigl(\mathcal{C}_t,\{(v_{j,k},r_{j,k},\mathcal{E}_{j,k})\}_{j,k}\bigr),\\
        \tau_t &\sim R_\theta(\cdot\mid x_t,y_t^A,y_t^B).
    \end{aligned}
\end{equation}
The task-conditioned rubric $\mathcal{C}_t=\{(c_k,w_k)\}_{k=1}^{K}$ is shared by both candidates. Each criterion $c_k$ has weight $w_k\geq0$, with $\sum_k w_k=1$.
For $j\in\{A,B\}$ and $k=1,\ldots,K$, $v_{j,k}$ is the score of candidate $j$ under $c_k$. The rationale $r_{j,k}$ explains this score, and $\mathcal{E}_{j,k}$ lists its evidence references to task requirements and available information.

Scores are measured on a comparable numerical scale across criteria, with higher values indicating better performance. We deterministically compute an overall score for each candidate through weighted aggregation and predict the preference by comparing the resulting scores:
\begin{equation}
    s_t^j=\sum_{k=1}^{K}w_kv_{j,k},\qquad
    \hat{\ell}_t=\arg\max_{j\in\{A,B\}}s_t^j.
\end{equation}

\section{Method}
\label{sec:method}

E-GenRM learns the evaluation chain defined in Section~\ref{sec:problem-formulation}. Section~\ref{sec:trajectory-data} converts real shopping trajectories into agent evaluation data. Section~\ref{sec:multi-stage-training} trains E-GenRM in three stages (Figure~\ref{fig:egenrm-framework}). 
Section~\ref{sec:tool_extension} introduces a separate retrieval extension that acquires e-commerce information when the current context is insufficient.
\subsection{From Shopping Trajectories to Agent Evaluation Data}
\label{sec:trajectory-data}

Training a shopping evaluator requires supervision about the requirements and trade-offs that drive a user's decision, but existing data do not directly provide such supervision. Platform behavior records what the user did, but rarely identifies why \cite{intenttune,lofa}. Agent-generated trajectories reflect the tasks and behaviors that existing agents already produce \cite{wang2026shoppingbench}, and their context is often incomplete. The central data problem is therefore to turn naturally occurring, heterogeneous behavior into agent evaluation data comprising complete decision contexts, candidate responses, and preference labels.

\textbf{Our central premise is that the evaluation standard of a shopping task emerges from the organization of behavior over time.} The sequence, repetition, and relation among events provide information that is unavailable from an isolated action. Such information reflects a task-specific intent that can change across sessions, rather than a fixed user profile. A purchase marks the end of a shopping task. Intermediate behavior before purchase (such as comparing products and querying product information) corresponds to decision steps at which the agent responds, and subsequent behavior indicates which response better fits the user. Accordingly, we trace each shopping trajectory backward from the purchase and construct the evaluation data in three steps.

\noindent\textbf{Trajectory Collection and Semantic Translation.} We collect more than 120 types of user behavior over one month and group platform events into shopping tasks with a consistent intent, each ending at a purchase decision. On average, a shopping trajectory spans 3.8 sessions and contains 162 behavioral events. Current-session behavior is restricted to the decision-time cutoff, and prior sessions provide historical context. We then translate the behavior into the interaction history defined in Section~\ref{sec:problem-formulation}. User-side behavior is rendered as user requests. Content returned by the platform is rendered as prior agent responses, with search and filtering represented as tool calls and their returned results. Of these behavior types, 126 serve as context and 18 are constructed into agent decision steps. Post-decision events and the purchase outcome are stored apart from the decision-time context and are used only to construct labels and check data quality.

\noindent\textbf{Decision-Step Extraction and Response Construction.} Tracing backward from the purchase, we decompose each shopping trajectory into multiple decision steps, yielding 4.6 steps per trajectory on average. The response type at each decision step is determined by the user behavior that triggers it. For example, when a user compares three products, we translate this behavior into a request to view their differences. The agent should present these differences, and its inferred preference should agree with the product the user later purchases. 

We take the response consistent with the user's subsequent behavior as the preferred response and construct the other candidate by perturbing it along a single dimension (such as omitting a historical preference, contradicting product facts, or recommending directly when clarification is needed). Each candidate pair therefore differs in a known aspect, and its preference label is determined jointly by the user's subsequent behavior and this controlled difference. 

At the final recommendation step, the candidate pair consists of the purchased product and an alternative considered in the same decision episode. The alternative shares the purchased product's leaf category and differs in an observable attribute. Tasks in which both products were purchased are excluded. Translation and response construction are performed by an LLM. We randomly sample 500 training decision steps for e-commerce annotators to review the translation and labels, yielding a pass rate of 94.2\%.

\noindent\textbf{Criteria Induction and Evaluation-Chain Construction.} We extend preference-conditioned evaluation-chain construction \cite{zhang2026pgenrm} to heterogeneous shopping behavior. At each decision step, a teacher examines the translated interaction history, together with candidate responses and product evidence. The teacher distinguishes explicit requirements from inferred preferences while reconciling redundant or conflicting signals. This synthesis yields a task-specific rubric $\mathcal{C}_t$ comprising weighted evaluation criteria. The teacher then evaluates both candidates under the rubric and produces criterion-level scores $v_{j,k}$, rationales $r_{j,k}$, and evidence references $\mathcal{E}_{j,k}$. We retain chains whose criteria are supported by the history, cover salient decision factors, and agree with the cited product facts, and whose preference agrees with the label. E-GenRM receives only the decision-time context and its available evidence. The filtered chains form $\mathcal{D}_{\mathrm{eval\_chain}}$ for training in Section~\ref{sec:multi-stage-training}.

\subsection{Multi-Stage Training}
\label{sec:multi-stage-training}

\subsubsection{\texorpdfstring{\textls[-45]{\textbf{Stage 1: Behavior-Grounded Supervised Fine-Tuning}}}{Stage 1: Behavior-Grounded Supervised Fine-Tuning}}

Stage 1 fine-tunes the model on $\mathcal{D}_{\mathrm{eval\_chain}}$ from Section~\ref{sec:trajectory-data}. Each example $(x_t,y_t^A,y_t^B,\tau_t)$ corresponds to a decision step. Its input is the decision context $x_t$ and the candidate pair defined in Section~\ref{sec:problem-formulation}, and its target is the filtered teacher evaluation chain $\tau_t$ from Section~\ref{sec:trajectory-data}. The model thus learns to derive the rubric from the visible context and then assess both candidates under each criterion. Current explicit requirements take precedence over conflicting historical inferences, keeping the learned rubric tied to the active task. Following next-token training for generative verification \cite{zhang2025generativeverifiers}, we supervise the complete evaluation chain $\tau_t$ and optimize
\begin{equation}
    \mathcal{L}_{\mathrm{SFT}}(\theta)=-\mathbb{E}_{(x_t,y_t^A,y_t^B,\tau_t)\sim\mathcal{D}_{\mathrm{eval\_chain}}}\left[\log p_\theta(\tau_t\mid x_t,y_t^A,y_t^B)\right].
\end{equation}

\subsubsection{\textbf{Stage 2: Criterion-Guided Reinforcement Learning}}

In Stage~1, E-GenRM learns to infer task-relevant criteria and express them as a rubric. Although SFT exposes the model to evidence-based demonstrations, self-generated assessments admit diverse reasoning paths and surface forms, so its token-level imitation objective only indirectly reflects whether these assessments are supported by evidence. This is particularly evident when high-level shopping requirements involve multiple product attributes. This motivates directly optimizing model-generated assessments against verifiable criterion-level feedback
\cite{yu2026shoppingcompanion}.

\noindent\textbf{Criterion-Level Factual Feedback.} We optimize complete evaluation chains with Group Relative Policy Optimization (GRPO) \cite{shao2024deepseekmath}, allowing the model to generate both the rubric and its candidate assessments. Training examples are drawn from decision steps in Section~\ref{sec:trajectory-data} whose users are disjoint from those in Stage 1. Each example contains only the decision context, the candidate pair, and the preference label, without a teacher evaluation chain. The reward combines final preference correctness with criterion-level feedback from a fact verifier. For each generated criterion, the verifier checks the factual claims in each candidate assessment against the visible evidence (such as product facts, returned tool results, and the interaction history). A claim citing an evidence reference absent from the visible input receives no factual credit, and a numerical comparison whose operands are addressable in the structured data is verified directly. This feedback evaluates the link from each criterion to the evidence used to assess it, providing partial credit for supported judgments within a chain. 

For criterion $k$, let $F_{j,k}\in\{0,0.5,1\}$ mark whether the assessment of candidate $j\in\{A,B\}$ is unsupported, partially supported, or supported by visible evidence. The criterion-level reward is
\begin{equation}
    r_k=\frac{F_{A,k}+F_{B,k}}{2},\qquad
    R_{\mathrm{crit}}=\frac{1}{K}\sum_{k=1}^{K}r_k.
\end{equation}
We combine the criterion-level signal with a reward for the final pairwise decision,
\begin{equation}
    R=R_{\mathrm{out}}+\lambda R_{\mathrm{crit}},
\end{equation}
where $R_{\mathrm{out}}$ rewards a correct final preference and $\lambda$ balances the two objectives. GRPO compares alternative evaluation chains for the same decision step, encouraging chains that make the right choice for the right evidence-based reasons.

\subsubsection{\textbf{Stage 3: Personalized Hard-Negative Training}}

As Stage 2 improves evidence-based judgment, the remaining errors concentrate on cases where both candidates satisfy the explicit shopping requirements. Verifying their price, functionality, and intended use then establishes only that both are suitable, and which one better fits this user depends on personal preferences supported by repeated interactions (such as material, style, or brand). Among the candidate pairs constructed in Section~\ref{sec:trajectory-data}, such cases mainly correspond to final recommendation steps that compare the purchased product with a considered alternative, and to perturbed candidates that alter a historical preference. We construct a targeted hard-negative set from these two types of candidate pairs for this decision boundary.

\noindent\textbf{Hard-Example Selection.} Hard-negative training has been used to refine personalized reward models \cite{zhang2026pgenrm}. We select hard examples using two complementary signals. The first signal is teacher disagreement. We retain pairs for which roughly half of the teachers' judgments disagree with the preference label, as these pairs expose ambiguous reasoning. The second signal is utility margin. We retain pairs with similar aggregated utilities, as these pairs lie near the decision boundary. All selected pairs satisfy explicit task constraints and are mixed with standard examples to preserve coverage while emphasizing fine-grained preference distinctions.

\noindent\textbf{Continued Training.} From the Stage 2 model, we continue training on this mixture with the same outcome and criterion-level objectives. This stage focuses learning on user-specific trade-offs while retaining the evidence-based judgment learned in Stage 2.

\subsection{E-GenRM-Bench Construction}
\label{sec:bench-construction}

E-GenRM-Bench evaluates whether a reward model can identify the agent response that better fits the user at a shopping decision step. Each evaluation item follows the interface in which a reward model evaluates shopping agent responses, comprising a decision-time context, two candidate responses, and a preference label. 

{\emergencystretch=1.5em
\noindent\textbf{Expert Annotation.}
Two e-commerce experts independently evaluate each candidate pair and determine the superior response. They are given the decision-time context, both responses, product evidence, and the user's subsequent behavior, allowing the annotation to reflect both response quality and the user's observed choice. Disagreements are resolved by a third expert; ties are excluded, and the adjudicated judgment serves as the reference label.
\par}

{\emergencystretch=1.5em\noindent\textbf{Evidence Separation and Evaluation Groups.} The construction pipeline separates decision evidence, preference labels, and reference criteria, with teacher-generated criteria retained as references for criterion-recall evaluation. For evaluation, a fixed manifest assigns each held-out decision step to one group. Decision steps containing behavior or leaf categories unseen during training are assigned to OOD. Among the remaining decision steps, those with high query or comparison complexity are assigned to Hard, and all others to Standard. The group assignments do not depend on evaluated-model predictions.\par}
\begin{table*}[!ht]
\centering
\caption{Comparison with LLM-as-a-Judge methods on E-GenRM-Bench (\%). \textbf{Bold}/\underline{underline}: best/second-best per column. Dark/light red: best/second-best per metric across Tables~\ref{tab:main-results} and~\ref{tab:main-results-2}. Marks exclude E-GenRM + retrieval. ---: not applicable.}
\label{tab:main-results}
\scriptsize
\setlength{\tabcolsep}{1.05pt}
\renewcommand{\arraystretch}{1.12}
\begin{tabular*}{\textwidth}{@{\extracolsep{\fill}}l*{20}{c}@{}}
\toprule
& \multicolumn{8}{c}{Open-source models} & \multicolumn{12}{c}{Proprietary models} \\
\cmidrule(lr){2-9}\cmidrule(lr){10-21}
Method & \multicolumn{4}{c}{Qwen3.5-9B} & \multicolumn{4}{c}{Qwen3.5-27B} & \multicolumn{4}{c}{GPT-5.6-Sol} & \multicolumn{4}{c}{Gemini-3.8-Flash} & \multicolumn{4}{c}{Qwen3.8-Max} \\
\cmidrule(lr){2-5}\cmidrule(lr){6-9}\cmidrule(lr){10-13}\cmidrule(lr){14-17}\cmidrule(lr){18-21}
 & Standard & Hard & OOD & Avg. & Standard & Hard & OOD & Avg. & Standard & Hard & OOD & Avg. & Standard & Hard & OOD & Avg. & Standard & Hard & OOD & Avg. \\
\midrule
\multicolumn{21}{l}{\emph{LLM-as-a-Judge}} \\
Direct & 56.49 & 54.55 & 48.21 & 54.96 & 62.21 & 45.45 & 57.14 & 58.98 & \underline{70.61} & 61.82 & \underline{69.64} & \underline{69.17} & 79.01 & \underline{67.27} & \cellcolor[rgb]{0.963,0.846,0.883}\underline{80.36} & \underline{77.48} & 58.40 & \underline{49.09} & 50.00 & 55.76 \\
CoT & 53.82 & 52.73 & 48.21 & 52.82 & 59.92 & 52.73 & 55.36 & 58.18 & \underline{70.61} & \underline{63.64} & 66.07 & 68.90 & \underline{80.15} & \cellcolor[rgb]{0.963,0.846,0.883}\textbf{69.09} & 73.21 & \underline{77.48} & 60.69 & \underline{49.09} & 55.36 & 58.18 \\
CheckEval~\cite{lee2024checkeval} & 51.15 & 41.82 & 50.00 & 49.60 & 53.05 & 47.27 & 50.00 & 51.74 & 54.58 & 50.91 & 55.36 & 54.16 & 68.70 & 65.45 & 71.43 & 68.63 & 58.40 & 41.82 & \textbf{58.93} & 56.03 \\
Dataset-Specific Rubric~\cite{wang2026dynamicrubrics} & 38.93 & 49.09 & 41.07 & 40.75 & 60.69 & 47.27 & \underline{58.93} & 58.45 & \textbf{76.72} & \cellcolor[rgb]{0.963,0.846,0.883}\textbf{69.09} & \textbf{73.21} & \textbf{75.07} & \cellcolor[rgb]{0.963,0.846,0.883}\textbf{82.82} & 65.45 & \cellcolor[rgb]{0.968,0.736,0.768}\textbf{83.93} & \cellcolor[rgb]{0.963,0.846,0.883}\textbf{80.43} & \underline{62.98} & \underline{49.09} & \underline{57.14} & \underline{60.05} \\
Instance-Specific Rubric~\cite{wang2026dynamicrubrics} & \underline{65.65} & \underline{63.64} & \underline{67.86} & \underline{65.68} & 56.87 & 38.18 & 57.14 & 54.16 & 54.20 & 56.36 & 55.36 & 54.69 & 55.34 & 52.73 & 58.93 & 55.50 & 56.87 & 47.27 & 51.79 & 54.69 \\
Rubrics on Trial~\cite{yang2026rubricstrial} & 56.11 & 54.55 & 48.21 & 54.69 & \underline{66.79} & \underline{56.36} & 53.57 & \underline{63.27} & 63.36 & \underline{63.64} & 53.57 & 61.93 & 55.34 & 50.91 & 64.29 & 56.03 & \textbf{63.36} & \textbf{56.36} & \textbf{58.93} & \textbf{61.66} \\
\midrule
\multicolumn{21}{l}{\emph{Ours}} \\
\egfirst{\textbf{E-GenRM}} & \cellcolor{egblue}\textbf{82.06} & \cellcolor{egblue}\textbf{67.27} & \cellcolor{egblue}\textbf{75.00} & \cellcolor{egblue}\textbf{78.82} & \cellcolor[rgb]{0.968,0.736,0.768}\textbf{85.88} & \cellcolor[rgb]{0.963,0.846,0.883}\textbf{69.09} & \cellcolor[rgb]{0.963,0.846,0.883}\textbf{80.36} & \cellcolor[rgb]{0.968,0.736,0.768}\textbf{82.57} & NOTE: &  &  &  &  &  &  &  &  &  &  &  \\
\egtightrows
\egfirst{{\tiny\quad $\Delta$ vs.\ same-scale best}} & \cellcolor{egblue}{\tiny\textcolor{red!80!black}{+3.82}} & \cellcolor{egblue}{\tiny\textcolor{green!45!black}{$-$3.64}} & \cellcolor{egblue}{\tiny\textcolor{green!45!black}{$-$3.57}} & \cellcolor{egblue}{\tiny\textcolor{red!80!black}{+1.88}} & \cellcolor{egblue}{\tiny\textcolor{red!80!black}{+17.18}} & \cellcolor{egblue}{\tiny\textcolor{red!80!black}{+12.73}} & \cellcolor{egblue}{\tiny\textcolor{red!80!black}{+12.50}} & \cellcolor{egblue}{\tiny\textcolor{red!80!black}{+16.35}} & \egnote{Compared to the best open-source baseline of comparable size ($\leq$10B for 9B; 27--32B for 27B), selected per metric.} \\
\egfirst{{\tiny\quad $\Delta$ vs.\ best baseline overall}} & \cellcolor{egblue}{\tiny\textcolor{green!45!black}{$-$0.76}} & \cellcolor{egblue}{\tiny\textcolor{green!45!black}{$-$3.64}} & \cellcolor{egblue}{\tiny\textcolor{green!45!black}{$-$8.93}} & \cellcolor{egblue}{\tiny\textcolor{green!45!black}{$-$1.61}} & \cellcolor{egblue}{\tiny\textcolor{red!80!black}{+3.06}} & \cellcolor{egblue}{\tiny\textcolor{green!45!black}{$-$1.82}} & \cellcolor{egblue}{\tiny\textcolor{green!45!black}{$-$3.57}} & \cellcolor{egblue}{\tiny\textcolor{red!80!black}{+2.14}} & \egnote{Compared to the best baseline among all methods in Tables~\ref{tab:main-results} and~\ref{tab:main-results-2}, selected per metric.} \\
\midrule
\egnormalrows
\egfirst{\textbf{E-GenRM + retrieval}} & \cellcolor{egblue}--- & \cellcolor{egblue}--- & \cellcolor{egblue}--- & \cellcolor{egblue}--- & \cellcolor{egblue}90.08 & \cellcolor{egblue}76.36 & \cellcolor{egblue}85.71 & \cellcolor{egblue}87.40 & \egnote{Retrieval is evaluated with Qwen3.5-27B (Section~\ref{sec:retrieval}).} \\
\egtightrows
\egfirst{{\tiny\quad $\Delta$ vs.\ best baseline overall}} & \cellcolor{egblue} & \cellcolor{egblue} & \cellcolor{egblue} & \cellcolor{egblue} & \cellcolor{egblue}{\tiny\textcolor{red!80!black}{+7.26}} & \cellcolor{egblue}{\tiny\textcolor{red!80!black}{+5.45}} & \cellcolor{egblue}{\tiny\textcolor{red!80!black}{+1.78}} & \cellcolor{egblue}{\tiny\textcolor{red!80!black}{+6.97}} & \egnote{Compared to the best baseline among all methods in Tables~\ref{tab:main-results} and~\ref{tab:main-results-2}, selected per metric.} \\
\bottomrule
\end{tabular*}
\end{table*}
\begin{table}[t]
\centering
\caption{Comparison with reward models on E-GenRM-Bench (\%). Notation follows Table~\ref{tab:main-results}.}
\label{tab:main-results-2}
\footnotesize
\setlength{\tabcolsep}{3pt}
\renewcommand{\arraystretch}{1.12}

\resizebox{\columnwidth}{!}{%
\begin{tabular}{@{}llcccc@{}}
\toprule
Method & Base model & Standard & Hard & OOD & Avg. \\
\midrule

Bradley--Terry RM~\cite{bradleyterry1952}
& Qwen3.5-9B
& 59.16
& 50.91
& 58.93
& 57.91 \\

Bradley--Terry RM
& Qwen3.5-27B
& 68.70	
& 52.73	
& 67.86	
& 66.22 \\

Skywork-Reward-V2~\cite{skyworkrewardv2}
& Qwen3-4B
& 75.57
& \cellcolor[rgb]{0.963,0.846,0.883}\underline{69.09}
& 67.86
& 73.46 \\

Skywork-Reward-V2
& Qwen3-8B
& 75.95
& 65.45
& \underline{78.57}
& 74.80 \\

Skywork-Reward-V2
& Llama-3.1-8B
& 78.24
& \cellcolor[rgb]{0.968,0.736,0.768}\textbf{70.91}
& 76.79
& 76.94 \\

RM-R1~\cite{rmr1}
& DeepSeek-Distilled-Qwen-7B
& 35.88
& 38.18
& 42.86
& 37.27 \\

RM-R1
& DeepSeek-Distilled-Qwen-32B
& 49.62
& 49.09
& 39.29
& 47.99 \\

RM-R1
& Qwen2.5-7B-Instruct
& 39.69
& 25.45
& 35.71
& 37.00 \\

RM-R1
& Qwen2.5-32B-Instruct
& 31.68
& 21.82
& 21.43
& 28.69 \\

GRAM-R$^2$~\cite{gramr2}
& Llama-3.1-8B
& 36.64
& 32.73
& 30.36
& 35.12 \\

RewardAnything~\cite{rewardanything}
& Qwen3-8B
& 51.15
& 54.55
& 44.64
& 50.67 \\

\midrule
\multicolumn{6}{@{}l}{\emph{Ours}} \\
\egfirst{\textbf{E-GenRM}} & \cellcolor{egblue}Qwen3.5-9B & \cellcolor{egblue}\underline{82.06} & \cellcolor{egblue}67.27 & \cellcolor{egblue}75.00 & \cellcolor{egblue}\underline{78.82} \\
\egfirst{\textbf{E-GenRM}} & \cellcolor{egblue}Qwen3.5-27B & \cellcolor[rgb]{0.968,0.736,0.768}\textbf{85.88} & \cellcolor[rgb]{0.963,0.846,0.883}\underline{69.09} & \cellcolor[rgb]{0.963,0.846,0.883}\textbf{80.36} & \cellcolor[rgb]{0.968,0.736,0.768}\textbf{82.57} \\
\midrule
\egfirst{\textbf{E-GenRM + retrieval}} & \cellcolor{egblue}Qwen3.5-27B & \cellcolor{egblue}90.08 & \cellcolor{egblue}76.36 & \cellcolor{egblue}85.71 & \cellcolor{egblue}87.40 \\

\bottomrule
\end{tabular}%
}
\end{table}
\subsection{Active E-Commerce Information Retrieval}
\label{sec:tool_extension}
The evaluation capability trained in the preceding stages relies only on the information available to the evaluator, whereas some e-commerce decisions depend on long-tail or rapidly changing facts (such as detailed compatibility, current inventory, or recent product feedback). These facts are often absent from the context and cannot be reliably recovered from model knowledge, which leaves the evaluator without decisive evidence. Retrieval can improve answers to e-commerce brand and category questions \cite{chen2025chineseecomqa}, and tool-augmented reward models likewise acquire external evidence before judging a response \cite{armthinker}. We therefore add an evidence-acquisition policy to E-GenRM while keeping its rubric generation and judgment procedure unchanged.

\noindent\textbf{Evidence Acquisition.} We extend the evaluation trajectory with retrieval actions and their observations, following the ReAct \cite{yao2023react} and adaptive evidence acquisition \cite{asai2024selfrag}. At each step, the evaluator can request product-encyclopedia or review evidence, or complete its judgment using the available context. Let $\mathcal{E}_{\mathrm{tool}}(\tau)$ collect the facts returned along trajectory $\tau$. These facts augment the initial evidence $e_t$ and support the same criterion-level assessments and preference aggregation defined in Section~\ref{sec:problem-formulation}.

\noindent\textbf{Retrieval-Need Supervision.} To supervise evidence acquisition, the extension data pairs two teacher evaluations of each task: one uses the original context, and the other additionally uses facts returned by the available tools. We assign a retrieval-need label $n=1$ when the original context lacks a decisive fact and the augmented context supplies evidence needed to establish or correct the preference. Tasks whose original evidence already supports the judgment receive $n=0$.

\noindent\textbf{Tool-Use Reward.} Let $c(\tau)\in\{0,1\}$ indicate whether trajectory $\tau$ invokes a tool. We retain the core evaluation reward $R_{\mathrm{out}}+\lambda R_{\mathrm{crit}}$ and align the retrieval decision with the need label through
\begin{equation}
R_{\mathrm{use}}(\tau)
=\begin{cases}
+1, & n=1,\ c(\tau)=1,\ R_{\mathrm{out}}(\tau)=1,\\
-1, & c(\tau)\ne n,\\
0, & \text{otherwise}.
\end{cases}
\label{eq:tool-use-reward}
\end{equation}
The positive reward couples a needed retrieval action with a correct final judgment. The combined objective is
\begin{equation}
R_{\mathrm{ext}}(\tau)=R_{\mathrm{out}}(\tau)
+\lambda R_{\mathrm{crit}}(\tau)+R_{\mathrm{use}}(\tau).
\label{eq:tool-reward}
\end{equation}
Continuing GRPO with this objective trains evidence acquisition alongside rubric-based preference evaluation.

\section{Experiments}
\label{sec:experiments}
\begingroup
\emergencystretch=2em

\subsection{Experimental Setup}
\label{sec:exp-setup}
\begin{table*}[tb]
\centering
\caption{Rubric generation on SRB (\%). Sem: semantic F1; RL: required-weighted ROUGE-L recall; F/H: Full/Hard. \textbf{Bold}/\underline{underline}: best/second-best per column. Dark/light red: best/second-best per metric. ---: not applicable.}
\label{tab:srb-results}
\scriptsize
\setlength{\tabcolsep}{1.5pt}
\renewcommand{\arraystretch}{1.12}
\begin{tabular*}{\textwidth}{@{\extracolsep{\fill}}l*{20}{c}@{}}
\toprule
& \multicolumn{8}{c}{Open-source models} & \multicolumn{12}{c}{Proprietary models} \\
\cmidrule(lr){2-9}\cmidrule(lr){10-21}
Method & \multicolumn{4}{c}{Qwen3.5-9B} & \multicolumn{4}{c}{Qwen3.5-27B} & \multicolumn{4}{c}{GPT-5.6-Sol} & \multicolumn{4}{c}{Gemini-3.8-Flash} & \multicolumn{4}{c}{Qwen3.8-Max} \\
\cmidrule(lr){2-5}\cmidrule(lr){6-9}\cmidrule(lr){10-13}\cmidrule(lr){14-17}\cmidrule(lr){18-21}
& Sem-F & Sem-H & RL-F & RL-H & Sem-F & Sem-H & RL-F & RL-H & Sem-F & Sem-H & RL-F & RL-H & Sem-F & Sem-H & RL-F & RL-H & Sem-F & Sem-H & RL-F & RL-H \\
\midrule
\multicolumn{21}{l}{\emph{Prompted rubric construction}} \\
Direct & 35.38 & 39.81 & 12.27 & 12.10 & 62.50 & 61.13 & 24.75 & 21.25 & 63.30 & 60.48 & \textbf{26.68} & \underline{23.33} & 67.19 & 65.14 & 24.17 & \textbf{22.51} & \textbf{68.84} & \underline{67.71} & 22.65 & 19.86 \\
CoT & 30.22 & 33.47 & 10.48 & 10.14 & 62.23 & 59.76 & 25.18 & 21.90 & \underline{66.35} & \underline{64.12} & 24.34 & 22.79 & 70.08 & 70.68 & 22.56 & 20.51 & \underline{68.40} & \textbf{67.77} & 23.70 & \underline{21.74} \\
Dataset-Specific Rubric~\cite{wang2026dynamicrubrics} & 50.56 & 48.41 & 15.66 & 14.45 & 53.79 & 51.54 & 15.05 & 13.59 & 53.69 & 51.49 & 14.87 & 14.27 & 65.22 & 62.90 & 13.62 & 12.85 & 53.62 & 51.40 & 15.71 & 14.87 \\
Instance-Specific Rubric~\cite{wang2026dynamicrubrics} & 48.37 & 51.68 & 16.72 & 15.65 & \underline{63.08} & \underline{61.98} & 25.47 & 21.64 & \textbf{67.03} & \textbf{64.81} & 24.29 & 22.86 & 71.40 & 70.99 & 22.99 & 21.63 & 63.68 & 62.67 & 22.30 & 20.21 \\
Rubrics on Trial~\cite{yang2026rubricstrial} & 18.44 & 22.71 & 6.66 & 7.10 & 60.16 & 58.13 & 24.48 & 21.27 & 66.11 & 63.89 & 24.43 & 22.62 & \cellcolor[rgb]{0.968,0.736,0.768}\textbf{72.78} & \cellcolor[rgb]{0.968,0.736,0.768}\textbf{73.39} & 24.30 & \underline{22.45} & 64.20 & 61.57 & 22.60 & 19.45 \\
\midrule
\multicolumn{21}{l}{\emph{Rubric generation from reward methods}} \\
RLCF~\cite{viswanathan2025rlcf} & \underline{54.22} & 54.57 & \underline{21.28} & 19.20 & 60.53 & 58.82 & 24.88 & 21.64 & 65.75 & 63.83 & 24.36 & 22.63 & \underline{72.48} & \underline{71.78} & \underline{24.40} & 21.89 & 65.12 & 61.90 & \underline{23.98} & 21.00 \\
CARMO~\cite{gupta2025carmo} & 52.55 & \underline{55.14} & 20.83 & \underline{19.49} & 61.05 & 59.70 & \underline{25.90} & \underline{22.75} & 64.55 & 62.28 & \underline{25.78} & \textbf{24.40} & 70.69 & 69.11 & \textbf{24.47} & 21.33 & 65.57 & 63.11 & \cellcolor[rgb]{0.963,0.846,0.883}\textbf{27.58} & \cellcolor[rgb]{0.963,0.846,0.883}\textbf{24.89} \\
\midrule
\egfirst{\textbf{E-GenRM}} & \cellcolor{egblue}\textbf{72.27} & \cellcolor{egblue}\textbf{71.69} & \cellcolor{egblue}\textbf{24.47} & \cellcolor{egblue}\textbf{21.33} & \cellcolor[rgb]{0.963,0.846,0.883}\textbf{72.53} & \cellcolor[rgb]{0.963,0.846,0.883}\textbf{72.56} & \cellcolor[rgb]{0.968,0.736,0.768}\textbf{28.67} & \cellcolor[rgb]{0.968,0.736,0.768}\textbf{25.38} & NOTE: &  &  &  &  &  &  &  &  &  &  &  \\
\egtightrows
\egfirst{{\tiny\quad $\Delta$ vs.\ same-scale best}} & \cellcolor{egblue}{\tiny\textcolor{red!80!black}{+18.05}} & \cellcolor{egblue}{\tiny\textcolor{red!80!black}{+16.55}} & \cellcolor{egblue}{\tiny\textcolor{red!80!black}{+3.19}} & \cellcolor{egblue}{\tiny\textcolor{red!80!black}{+1.84}} & \cellcolor{egblue}{\tiny\textcolor{red!80!black}{+9.45}} & \cellcolor{egblue}{\tiny\textcolor{red!80!black}{+10.58}} & \cellcolor{egblue}{\tiny\textcolor{red!80!black}{+2.77}} & \cellcolor{egblue}{\tiny\textcolor{red!80!black}{+2.63}} & \egnote{Compared to the best baseline of comparable size ($\leq$10B for 9B; 27--32B for 27B), selected per metric.} \\
\egfirst{{\tiny\quad $\Delta$ vs.\ best baseline overall}} & \cellcolor{egblue}{\tiny\textcolor{green!45!black}{$-$0.51}} & \cellcolor{egblue}{\tiny\textcolor{green!45!black}{$-$1.70}} & \cellcolor{egblue}{\tiny\textcolor{green!45!black}{$-$3.11}} & \cellcolor{egblue}{\tiny\textcolor{green!45!black}{$-$3.56}} & \cellcolor{egblue}{\tiny\textcolor{green!45!black}{$-$0.25}} & \cellcolor{egblue}{\tiny\textcolor{green!45!black}{$-$0.83}} & \cellcolor{egblue}{\tiny\textcolor{red!80!black}{+1.09}} & \cellcolor{egblue}{\tiny\textcolor{red!80!black}{+0.49}} & \egnote{Compared to the best baseline among all methods, selected per metric.} \\
\bottomrule
\end{tabular*}
\end{table*}
\textbf{E-GenRM-Bench.} Following the construction in Section~\ref{sec:bench-construction}, E-GenRM-Bench contains 373 expert-annotated decision steps with 2,387 teacher-generated reference criteria and decision-time contexts collectively containing 45,133 behavior events that may inform reward judgments. For expert annotation, at most two decision steps are sampled from each trajectory. The selected steps cover 231 users and 264 product categories and form three mutually exclusive groups, Standard (262), Hard (55), and OOD (56).

\textbf{Shopping Reasoning Bench.} SRB contains 525 expert-authored shopping missions and 10,863 expert rubric items~\cite{fan2026shoppingreasoning}. Each method generates criteria from the current and preceding user messages, without access to the expert rubrics used for scoring. We report the Full set and the official Hard subset of 108 missions included in it.

\textbf{Metrics.} E-GenRM-Bench accuracy (Acc.) measures agreement between model judgments and expert preference labels. A decision step is counted as correct only if the model selects the preferred response under both candidate orderings; order-inconsistent or invalid outputs are counted as incorrect. Avg.\ denotes micro-averaged accuracy over all decision steps. For ablations, lexical reference-criterion recall (RR-L) measures coverage of teacher-generated reference criteria and is macro-averaged over decision steps. On SRB, semantic F1 (Sem-F1) scores generated--reference criterion pairs with BERTScore~\cite{zhang2020bertscore} and computes F1 from their maximum-weight one-to-one matching. We adapt ROUGE-L~\cite{lin2004rouge} to compute required-weighted recall (RL), which measures lexical recovery with greater weight on required items. SRB scores are averaged first over turns within each mission and then over missions.

\textbf{Baselines.} On E-GenRM-Bench, we compare E-GenRM with LLM-as-a-Judge methods and reward models. LLM-as-a-Judge baselines are evaluated on five base models, including two open-source models (Qwen3.5-9B and Qwen3.5-27B) and three proprietary models (GPT-5.6-Sol, Gemini-3.8-Flash, and Qwen3.8-Max). Direct produces the output directly, whereas CoT~\cite{kojima2022zeroshot} reasons explicitly before producing it. CheckEval~\cite{lee2024checkeval} uses a checklist. Dataset-Specific Rubric~\cite{wang2026dynamicrubrics} constructs criteria shared across the dataset, Instance-Specific Rubric~\cite{wang2026dynamicrubrics} constructs task-specific criteria, and Rubrics on Trial~\cite{yang2026rubricstrial} refines criteria through synthetic comparisons. Scalar reward models include Bradley--Terry RM~\cite{bradleyterry1952}, trained on the same shopping preference pairs, and released Skywork-Reward-V2 checkpoints~\cite{skyworkrewardv2}. Generative reward models include RM-R1~\cite{rmr1} and GRAM-R$^2$~\cite{gramr2}, which generate reasoning-based judgments, and RewardAnything~\cite{rewardanything}, which evaluates candidates against supplied principles. On SRB, the LLM-as-a-Judge baselines generate rubrics instead of preference judgments, and we additionally compare the query-conditioned rubric-generation components of RLCF~\cite{viswanathan2025rlcf} and CARMO~\cite{gupta2025carmo}. We evaluate released reward checkpoints through their preference-judgment interfaces and restrict the SRB comparison to methods adapted for standalone rubric generation.

\textbf{Implementation Details.} We train Qwen3.5-9B and Qwen3.5-27B following Section~\ref{sec:multi-stage-training}. SFT uses LoRA~\cite{hu2022lora} with rank 64 and scaling factor 128 for 1 epoch, a learning rate of $5\times10^{-5}$, a global batch size of 32, and 3\% warmup. RL updates all parameters at $10^{-6}$ with an optimizer batch size of 16. Each RL rollout batch contains 16 prompts with four responses per prompt. The three stages use 4,492, 1,503, and 747 examples, respectively, and Stage 3 includes 135 hard examples. Training and evaluation users are disjoint. GPT-5.6-Sol serves as the teacher that generates evaluation chains and as the fact verifier that provides criterion feedback. We set the factual-feedback weight to $\lambda=0.25$.
\subsection{Main Results}
\label{sec:main-results}

Tables~\ref{tab:main-results} and~\ref{tab:main-results-2} report preference accuracy on E-GenRM-Bench against LLM-as-a-Judge methods and reward models, respectively, and Table~\ref{tab:srb-results} evaluates rubric generation on SRB, an external benchmark held out from training and model selection. In Avg.\ Acc., E-GenRM outperforms all same-scale baselines at both scales, and E-GenRM-27B further surpasses every proprietary-model configuration. 
Specifically, E-GenRM-27B attains 82.57\%, exceeding the strongest baseline, Gemini-3.8-Flash with Dataset-Specific Rubric, by 2.14 percentage points. 
E-GenRM-9B attains 78.82\%, exceeding the best same-scale open-source baseline, Skywork-Reward-V2-Llama-3.1-8B, by 1.88 points. With active e-commerce information retrieval (Section~\ref{sec:retrieval}), E-GenRM-27B reaches 87.40\%, widening its lead over the strongest baseline to 6.97 points.

\textbf{Comparison with LLM-as-a-Judge Methods.}
Under the same base model, E-GenRM outperforms every prompting baseline on all subsets. Relative to the strongest prompting baseline, E-GenRM-9B and E-GenRM-27B improve Avg.\ Acc.\ by 13.14 and 19.30 points, respectively. On the Hard subset, where a decision step requires aggregating multiple queries or comparing multiple product attributes, E-GenRM-27B gains 12.73 points. The largest gain of E-GenRM-27B emerges on the OOD subset, where it reaches 80.36\% versus 58.93\% for the strongest prompting baseline, Dataset-Specific Rubric, a margin of 21.43 points. Whereas Dataset-Specific Rubric applies a single rubric across the entire dataset, E-GenRM generates a rubric conditioned on the context of each decision step, which is consistent with its gains on unseen behavior and product categories.

\textbf{Comparison with Reward Models.}
E-GenRM-9B and E-GenRM-27B achieve higher Avg.\ Acc.\ than every reward model in Table~\ref{tab:main-results-2}, and E-GenRM-27B also leads on Standard and OOD. The Bradley--Terry RM shares the shopping preference data and base model with E-GenRM-27B but outputs only a scalar score, whereas E-GenRM first generates task-specific criteria and then evaluates the candidates against each criterion. Under this controlled comparison, E-GenRM-27B exceeds it by 16.35 points. 
For generative reward models, E-GenRM-9B exceeds the strongest one.
These results suggest shopping evaluation calls for training data and objectives tailored to e-commerce challenges, such as inducing evaluation criteria from real shopping behavior and verifying candidate claims on product attributes against facts.

\textbf{Rubric Generation on SRB.}
The rubric generation ability of E-GenRM transfers to the external benchmark. With either base model, E-GenRM outperforms all methods built on the same base model across all four metrics. Because SRB criteria are generated solely from user messages, this result indicates that the rubric generation learned in Stage~1 is not tied to the behavioral inputs of E-GenRM-Bench. Across all methods, E-GenRM-27B achieves the highest lexical recall, outperforming the strongest
baselines by 1.09 points on Full and 0.49 on Hard, while trailing the best proprietary models in semantic F1 by only 0.25 points on Full and 0.83 on Hard.

\begin{table}[tb]
\centering
\caption{Qwen3.5-9B ablations on E-GenRM-Bench (\%).
Each entry reports Acc./RR-L.
Blocks share an initialization; w/o removes the named component.
--: unavailable RR-L.}
\label{tab:ablation}
\scriptsize
\setlength{\tabcolsep}{2.1pt}
\renewcommand{\arraystretch}{1.12}

\resizebox{\columnwidth}{!}{%
\begin{tabular}{@{}lcccc@{}}
\toprule
Variant
& Standard
& Hard
& OOD
& Avg. \\
& Acc./RR-L
& Acc./RR-L
& Acc./RR-L
& Acc./RR-L \\
\midrule

\multicolumn{5}{@{}l}{
  \textbf{Stage 1: Behavior-grounded SFT}
  \quad\emph{(init.: Qwen3.5-9B)}
} \\

\textbf{Evaluation-chain SFT}
& \textbf{70.61/20.26}
& \textbf{61.82/18.32}
& \textbf{66.07/20.83}
& \textbf{68.63/20.05} \\

\quad w/o evaluation chain (label-only SFT)
& 66.79/---
& 56.36/---
& 53.57/---
& 63.27/--- \\

\midrule

\multicolumn{5}{@{}l}{
  \textbf{Stage 2: Criterion-guided RL}
  \quad\emph{(init.: Stage 1 model)}
} \\

\textbf{Criterion-guided GRPO}
& \textbf{75.95/27.79}
& \textbf{65.45/24.06}
& \textbf{67.86/24.33}
& \textbf{73.19/26.74} \\

\quad w/o GRPO (continued SFT)
& 72.90/24.90
& 63.64/23.51
& 66.07/22.17
& 70.51/24.30 \\

\quad w/o outcome reward
& 60.31/12.97
& 40.00/12.11
& 39.29/8.67
& 54.16/12.23 \\

\quad w/o factual feedback
& 72.14/22.05
& 65.45/24.45
& 66.07/15.67
& 70.24/21.51 \\

\midrule

\multicolumn{5}{@{}l}{
  \textbf{Stage 3: Personalized hard-negative training}
  \quad\emph{(init.: Stage 2 model)}
} \\

\textbf{(Ours) E-GenRM}
& \textbf{82.06/15.45}
& \textbf{67.27/13.84}
& \textbf{75.00/17.00}
& \textbf{78.82/15.42} \\

\quad w/o personalized hard negatives
& 77.86/21.60
& 61.82/21.15
& 66.07/22.67
& 73.73/21.69 \\

\bottomrule
\end{tabular}%
}
\end{table}
\subsection{Ablation Studies}
\label{sec:ablation}
Table~\ref{tab:ablation} ablates the training procedure on Qwen3.5-9B by stage.

\textbf{Behavior-Grounded SFT.}
Compared with label-only SFT, evaluation-chain supervision improves Avg.\ Acc.\ by 5.36 points. The gain is largest on OOD (12.50 points) and smallest on Standard (3.82 points), which is consistent with evaluation chains teaching the model to derive criteria from the visible context rather than merely fit preference labels.

\textbf{Criterion-Guided Reinforcement Learning.}
Criterion-guided GRPO outperforms continued SFT by 2.68 points. To reward components, the outcome reward has the larger effect. Removing it reduces Avg.\ Acc.\ by 19.03 points, with drops of 25.45 and 28.57 points on Hard and OOD. Removing factual feedback reduces Avg.\ Acc.\ by 2.95 points and Avg.\ RR-L by 5.23 points, including an 8.66-point RR-L drop on OOD. Factual feedback therefore affects criterion recall more than accuracy, which is consistent with the fact verifier providing feedback at the criterion level.

\textbf{Personalized Hard-Negative Training.}
Compared with ordinary continued RL from the same Stage~2 model, personalized hard-negative training improves Avg.\ Acc.\ by 5.09 points and OOD accuracy by 8.93 points, while Avg.\ RR-L decreases by 6.27 points. In these hard negatives, both candidates satisfy the explicit requirements and must be distinguished by personal preferences. This is consistent with the model exploring evaluation criteria beyond the teacher references to improve accuracy.

\subsection{Effect of Active E-Commerce Information Retrieval}
\label{sec:retrieval}
Table~\ref{tab:retrieval} evaluates active e-commerce information retrieval of our best-performing variant, E-GenRM-27B, using product-encyclopedia search and review search as tools. We compare E-GenRM-27B with and without tool-use training, each evaluated without retrieval and with autonomous retrieval. The autonomous conditions share the same retrieval instruction and a one-call budget. As a reference, fixed retrieval supplies the tool-trained model with the results of two programmatic calls before judgment.

\begin{table}[tb]
\centering
\scriptsize
\caption{Active e-commerce information retrieval of E-GenRM-27B on E-GenRM-Bench (\%). $B$ is the call budget, and a view is one candidate order. Bold marks the better accuracy within each matched pair.}
\label{tab:retrieval}

\setlength{\tabcolsep}{1.5pt}
\renewcommand{\arraystretch}{1.05}

\begin{tabular*}{\columnwidth}{
  @{}
  l@{\hspace{2pt}}
  c@{\hspace{2pt}}
  c@{\hspace{5pt}\extracolsep{\fill}}
  cccccc
  @{}
}
\toprule
\multicolumn{3}{c}{Configuration}
& \multicolumn{4}{c}{Accuracy (\%) $\uparrow$}
& \multicolumn{2}{c}{Tool usage} \\
\cmidrule(lr){1-3}
\cmidrule(lr){4-7}
\cmidrule(l){8-9}

Retrieval
& \makecell{Tool\\training}
& $B$
& Standard
& Hard
& OOD
& Avg.
& \makecell{Call rate\\(\%)}
& \makecell{Calls\\per view} \\
\midrule

None & No & 0
& 85.88
& 69.09
& \textbf{80.36}
& 82.57
& 0.00
& 0.000 \\

None & Yes & 0
& \textbf{86.26}
& \textbf{72.73}
& \textbf{80.36}
& \textbf{83.38}
& 0.00
& 0.000 \\

\midrule

Autonomous & No & 1
& 88.55
& \textbf{80.00}
& 82.14
& 86.33
& 95.84
& 0.958 \\

Autonomous & Yes & 1
& \textbf{90.08}
& 76.36
& \textbf{85.71}
& \textbf{87.40}
& 22.12
& 0.221 \\

\midrule

Fixed & Yes & 2
& 90.46
& 76.36
& 82.14
& 87.13
& 100.00
& 2.000 \\
\bottomrule
\end{tabular*}
\end{table}
\textbf{Effect of Retrieval.}
Autonomous retrieval improves Avg.\ Acc.\ by 3.76 points for E-GenRM-27B and by 4.02 points for the tool-trained model. This is consistent with the motivation in Section~\ref{sec:tool_extension} that some shopping decisions depend on e-commerce information absent from the context.

\textbf{Effect of Tool-Use Training.}
Under autonomous retrieval, tool-use training improves Avg.\ Acc.\ by 1.07 points and lowers the call rate to 22.12\% (95.84\% without training). This is consistent with $R_{\mathrm{use}}$ penalizing calls when the context is sufficient. Compared with fixed retrieval, the tool-trained autonomous policy achieves comparable accuracy (87.40\% vs.\ 87.13\%) with 0.221 calls per view.

\begin{table}[tb]
\centering\footnotesize
\caption{Offline evaluation of the Shopping Agent post-trained with E-GenRM-27B or the Expert Judge (expert-calibrated Gemini-3.8-Flash). Prod., Quality, and Pers. are sub-scores, and Overall is their mean.}
\label{tab:policy-training}
\setlength{\tabcolsep}{3pt}
\begin{tabular*}{\columnwidth}{@{\extracolsep{\fill}}llccc@{\hspace{10pt}}c@{}}
\toprule
\multirow{2}{*}{Training} & \multirow{2}{*}{Evaluator} & \multicolumn{3}{c}{Sub-scores $\uparrow$} & \multirow{2}{*}{\textbf{Overall} $\uparrow$} \\
\cmidrule(lr){3-5}
 & & Prod. & Quality & Pers. & \\
\midrule
None & --- & 0.693 & 0.781 & 0.486 & 0.653 \\
\midrule
\multirow{2}{*}{GRPO} & Expert Judge & 0.759 & 0.858 & 0.528 & 0.715 \\
 & E-GenRM-27B & \textbf{0.777} & \textbf{0.879} & \textbf{0.529} & \textbf{0.728} \\
\midrule
\multirow{2}{*}{DPO} & Expert Judge & \textbf{0.760} & 0.806 & 0.493 & 0.686 \\
 & E-GenRM-27B & 0.747 & \textbf{0.864} & \textbf{0.505} & \textbf{0.705} \\
\bottomrule
\end{tabular*}
\end{table}

\begin{table}[tb]
\centering
\caption{Online A/B test results of Shopping Agent over 14 days. $\Delta$ is the change relative to the control; rates use percentage points (pp), and P99 latency uses relative change.}
\label{tab:online-results}
\begin{tabularx}{0.9\columnwidth}{
  @{}
  >{\raggedright\arraybackslash}X
  >{\centering\arraybackslash}c
  >{\centering\arraybackslash}c
  @{}
}
\toprule
Metric & $\Delta$ & $p$-value \\
\midrule
Effective assistance rate & +1.39 pp  & 0.0433 \\
Satisfaction rate         & +2.33 pp  & 0.0082 \\
Card coverage             & +7.47 pp  & 0.0067 \\
P99 latency $\downarrow$  & $-17.88\%$ & --- \\
\bottomrule
\end{tabularx}
\end{table}

\subsection{Application to a Deployed Shopping Agent}
\label{sec:policy-training}

We use E-GenRM-27B to post-train the Taobao Omni-Shop Agent, which serves
over 100,000 user queries daily. We
apply DPO~\cite{rafailov2023dpo} and GRPO~\cite{shao2024deepseekmath},
comparing E-GenRM-27B supervision with an expert-calibrated
Gemini-3.8-Flash Judge. For DPO, each evaluator ranks candidate responses
to construct preferred--rejected pairs; for GRPO, it evaluates sampled
responses to provide reward signals. Within each algorithm, we hold the
policy initialization, task pool, and update budget fixed across
supervision sources, and include the policy before post-training as a
baseline.

\textbf{Offline Evaluation.}
A proprietary e-commerce evaluation model scores product suitability
(Prod.), response quality (Quality), and personalized preference alignment
(Pers.); Overall denotes their arithmetic mean
(Table~\ref{tab:policy-training}). E-GenRM supervision improves Overall score by
0.075 under GRPO and 0.052 under DPO, exceeding the expert Judge by 0.013
and 0.019. The most pronounced improvement is in Quality,
where E-GenRM leads the expert Judge by 0.021 and 0.058.
These gains are consistent with E-GenRM's evaluation chain, which provides policy training with shopping-specific supervision that reflects both task requirements and personalized trade-offs.

\textbf{Online A/B Results.}
In a 14-day A/B test on a Taobao Shopping Agent, the agent post-trained
with E-GenRM significantly improves effective shopping assistance, user
satisfaction, and product-card coverage, and reduces P99 latency by
17.88\% (Table~\ref{tab:online-results}).

\par\endgroup

\section{Conclusion}
\label{sec:conclusion}

In this paper, we presented E-GenRM, an integrated data and modeling framework for e-commerce reward modeling in support of open-ended shopping-agent post-training. Its central insight is to derive evaluation criteria from naturally occurring shopping behavior and translate these diverse real-world behavioral patterns into shopping-agent trajectories, thereby providing more open-ended training data for reward models robust to diverse agent behaviors. E-GenRM constructs agent evaluation data and E-GenRM-Bench from real shopping trajectories, learns evidence-grounded evaluation through three-stage training, and acquires missing e-commerce information through an active e-commerce information retrieval extension. we offered a reference for reward design and agent post-training in related e-commerce applications. Experiments show that E-GenRM outperforms all same-scale baselines on E-GenRM-Bench and generalizes to unseen behavior, product categories, and an external benchmark. In an online A/B test, the shopping agent post-trained with E-GenRM improves the effective shopping-assistance rate and user satisfaction.

\clearpage 
\bibliographystyle{unsrt}
\bibliography{references}

\end{document}